\documentclass{aa}
\def\rxj{RX J1131-1231}

\usepackage{natbib}
\usepackage{graphicx}
\usepackage{amssymb}

\begin{document}
\title{Test of relativistic gravity using microlensing of relativistically broadened lines in gravitationally lensed quasars}
\titlerunning{Relativistic gravity test with microlensing of spectral lines}
\author{A. Neronov $^{1}$, Ie.Vovk$^2$}
\institute{
1. ISDC, Astronomy Department, University of Geneva, Ch. d'Ecogia 16, 1290, Versoix, Switzerland \\
2. Max Planck Institut f\"ur Physik, F\"ohringer ring 6, 80805, Munich, Germany 
}

\begin{abstract}
{}
{We show that observation of the time-dependent effect of microlensing of relativistically broadened emission lines (such as e.g. the Fe K$\alpha$ line in X-rays) in strongly lensed quasars could provide data on  celestial mechanics of circular orbits in the direct vicinity of the horizon of supermassive black holes.}
{This information can be extracted from the observation of evolution of  red / blue edge of the magnified line just before and just after the period of crossing of the innermost stable circular orbit by the microlensing caustic. The functional form of this evolution is insensitive to numerous astrophysical parameters of the accreting black hole and of the microlensing caustics network system (as opposed to the evolution the full line spectrum).}
{
Measurement of the temporal evolution of the red / blue edge could provide a precision measurement of the radial dependence of the gravitational redshift and of velocity of the circular orbits, down to the innermost stable circular orbit. These measurements could be used to discriminate between the General Relativity and alternative models of the relativistic gravity in which the dynamics of photons and massive bodies orbiting the gravitating centre is different from that of the geodesics in the Schwarzschild or Kerr space-times. 
}
{}
\end{abstract}


\maketitle

\section{Introduction}

General Relativity (GR) is by far the most popular theory of the relativistic gravity, tested with high precision via a set of ``classical'' tests~\citep{will14} in its weak field regime, when the gravitational potential is
\begin{equation}
  U=\frac{G_N M}{R}\simeq 10^{-6}\left[\frac{M}{M_\odot}\right]\left[\frac{R}{R_\odot}\right]^{-1}\ll 1,
\end{equation}
where $M$ and $R$ are the mass and size of the gravitating system (we use the system of units in which the speed of light $c=1$). However, none of the classical tests of GR provides a verification of the theory in the strong-field regime $U \sim 1$. 

At the same time, several observational phenomena occur in this regime. The $U \sim 1$ regime of gravity is influencing the physical processes in the vicinity of neutron stars and stellar mass black holes in pulsars, in the X-ray binaries and in the cores of massive stars exploding as supernovae. Electromagnetic emission from supermassive black holes in the centre of the Milky Way and other galaxies is also affected by the strong gravity effects. Finally, the dynamics of expansion of the Universe within the Hubble distance $R_H$ occurs in the strong field regime with $U\sim G_N \rho_{0} R_H^2\sim 1$ (here $\rho_0\simeq 10^{-29}$~g/cm$^3$ is the critical density of the Universe). 

Although the relativistic gravity effects certainly influence the observational properties of the sources powered by neutron stars and black holes~\citep{fabian00,remillard06,psaltis08}, large uncertainty of the physical conditions and parameters describing physical phenomena in these systems does not allow to test the GR. The unique nature of the expanding Universe system also precludes the possibility to test the theory.  The dynamical models of the Universe do not agree with observations in a straightforward way, as it is clear from the necessity of introduction of the Dark Matter and Dark Energy in certain fine-tuned proportions $\Omega_m\simeq 0.3, \Omega_\Lambda\simeq 0.7$~\citep{planck15} into the description of the dynamics of the Universe. 

GR is not the unique viable theoretical model of relativistic gravity (see e.g.~\cite{Clifton12,Berti15} for reviews). Some of the alternative theories are motivated by the possibility to solve the problem of the Dark Matter and Dark Energy. Others are introduced to achieve compatibility of the relativistic gravity with quantum mechanics. All the extensions and modifications of GR have the same weak field limit as the GR and as such are compatible with the classical tests of GR. At the same time, predictions of the extensions / modifications of GR in the strong field regime could strongly deviate from the predictions of the GR. In particular, the dynamics of the Universe and the properties of the black holes in different alternative models of relativistic gravity could be significantly different.

Perhaps, the most straightforward possibility to test  different properties of black holes in the alternative models of relativistic gravity is via a measurement of the laws of celestial mechanics in the strong gravitational field. After all, it is the study of the celestial mechanics of planets in the Solar system that led to the formulation and verification of the non-relativistic theory of gravity in the past.

\begin{figure}
  \includegraphics[width=\linewidth]{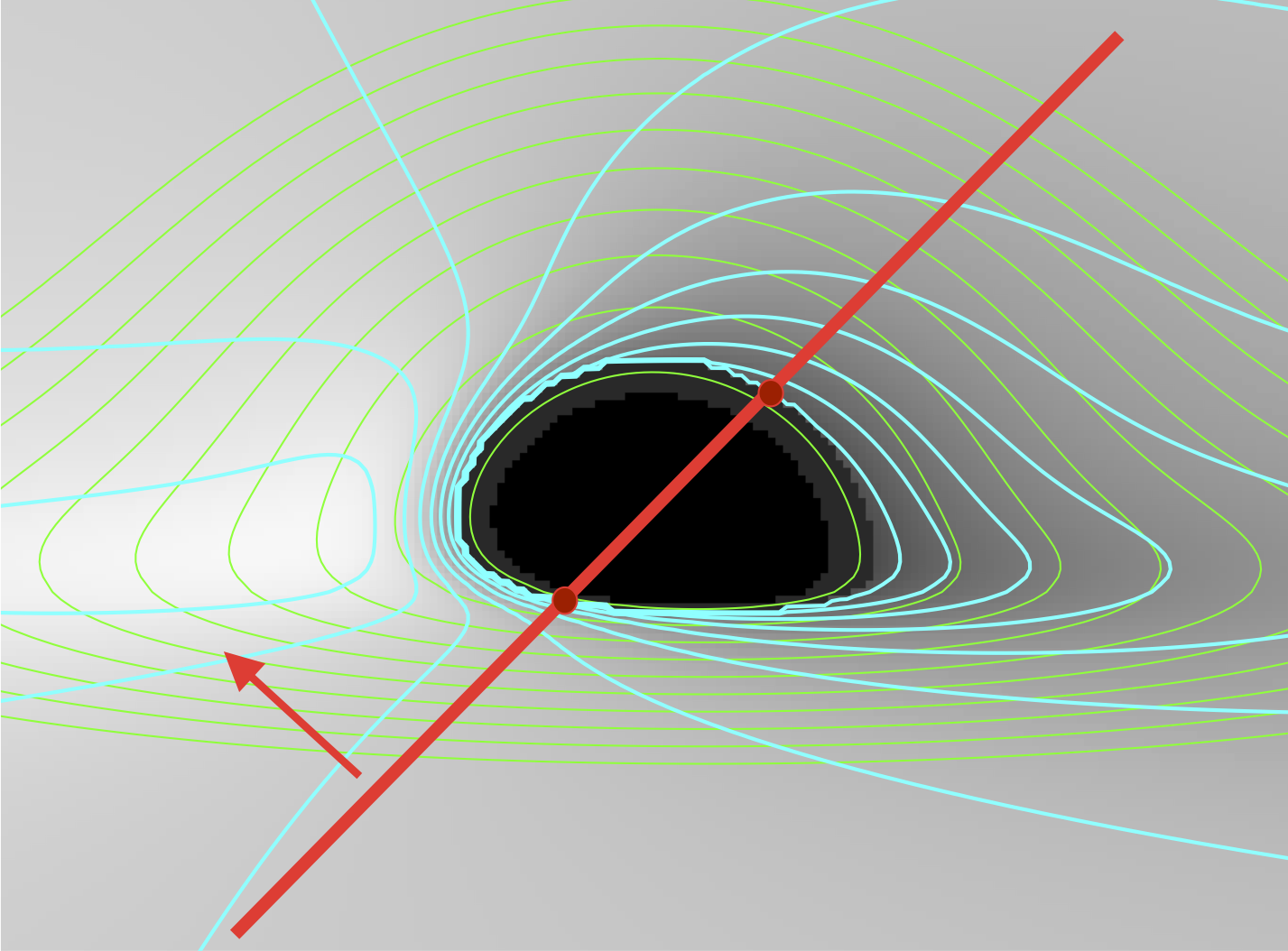}
  \caption{Map of the red/blue shift of an emission line from the equatorial disk around black hole with $a=0.9$ inclined at $80^\circ$  with respect to the line of sight. Green thin contours show the radial coordinate of the point of emission from the black hole in the Boyer -Lindquist coordinates. Contours start from $r=2$ with the increment of $\Delta r=2$ up to $r=20$.  Cyan thick contours show the redshift factor, starting from 0.3 with increments of 0.1.   The trajectory of caustics inclined with respect to the line of nodes of the disk is shown by the red line.}
  \label{fig:redshift_image}
\end{figure}

Several possibilities exist for such measurements using the observations of the supermassive black hole in the centre of the Milky Way.  Observations in the millimetre wavelength range have reached the angular resolution of about $10\ \mu$as necessary to  resolve the event horizon of the supermassive black hole in  the Galactic Centre and several nearby galaxies~\citep{Doeleman08,Doeleman12}. Precision of the astrometric measurements with the infra-red interferometric observations will also reach a comparable angular scale with the GRAVITY instrument at VLT~\citep{Gliessen10}. 

The angular size scales of more distant supermassive black holes in the galaxies beyond the Milky Way are typically too small to be directly resolvable. For these distant black holes a potential possibility  for the study of strong gravity effects is via the measurement of the Doppler and gravitational redshift of atomic lines emitted from the innermost portion of the accretion disks~\citep{fabian00}. However, if such measurements are not complemented by the information on spatial location of the portions of the disk with certain red / blue shifts, no real study of ``celestial mechanics'' of the orbits around the black hole is possible. 

In what follows we propose a method for the spatial localization of the red / blue shifted parts of the accretion disks. The method is based on the observation of the effect of microlensing of the emission from strongly gravitationally lensed quasars~\citep{dai10,chen12,chartas12,macleod15}. The microlensing caustics, passing in front of the accretion disk, work as a ``scanning magnifying glass'', helping to resolve the region of emission of the relativistically broadened lines. We argue that measurement of the time evolution of the red / blue edges of the line, magnified by the caustics, provides the information on the radial dependence of the rotation velocity and gravitational redshift down to the location of the innermost stable circular orbit (ISCO) around the black hole. This opens a possibility to test of the GR in the strong field regime.

\section{Relativistically broadened lines from the black hole accretion disks}

Radiatively efficient accretion onto supermassive black holes in quasars and Seyfert galaxies is conventionally assumed to proceed through an accretion disk which rotates at nearly Keplerian velocities in the equatorial plane around a black hole~\citep{sunyaev73,novikov74}. Photons emitted by the matter moving along the circular orbits are gravitationally redshifted and also red / blue shifted by the Doppler effect. The red and blue shifts depend on the position of the emission point and on the direction of photon propagation. Superposition of the red/blue shifted atomic line emission from different positions in the accretion disk leads to the occurrence of relativistically broadened emission lines~\citep{tanaka95,fabian00}. 

Calculation of the spectral shape of the broad line can be done following the standard procedure of ``ray shooting'' from the observer towards the black hole~\citep{bromley96,fanton97}. Fig.~\ref{fig:redshift_image} shows an example of such calculation for a particular case of the black hole with the specific angular momentum $a=0.9G_NM$ ($M$ is the black hole mass). The image is calculated by tracing the trajectories of photons emitted along the ``line of sight'', inclined at $i=80^\circ$  with respect to the black hole rotation axis. Photon trajectories start at a remote ``screen'', placed perpendicularly to the line of sight. Photons are back-traced from the ``screen'' till the moment when their trajectories intersect the equatorial plane, which contains the accretion disk, or when they hit the black hole horizon. Photons crossing the equatorial plane ${\cal N}$ times form the ${\cal N}$th order image (only the first order image is shown in Fig. \ref{fig:redshift_image}). Photon trajectories are calculated using the Runge-Kutta method for the solution of the first-order differential equations of the null geodesics~\citep{bardeen}. Only photons which hit the equatorial plane at the distances larger than the distance of the ISCO~\citep{bardeen} are contributing to the image.

The rotation velocity of the prograde circular orbit at the distance $r$ in the equatorial plane around the black hole is
\begin{equation}
  V^{(\phi)}=\frac{\sqrt{R_g}(r^2-2aR_g^{1/2}r^{1/2}+a^2)}{\sqrt{\Delta}(r^{3/2}+aR_g^{1/2})}
\end{equation}
in the Locally Non-Rotating Reference Frame (LNRF), defined by the tetrad
\begin{eqnarray}
  \hat e^{(0)}&=&\sqrt{\frac{\Sigma\Delta}{A}}dt\nonumber\\
  \hat e^{(1)}&=&\sqrt{\frac{\Sigma}{\Delta}}dr\nonumber\\
  \hat e^{(2)}&=&\sqrt{\Sigma}d\theta\nonumber\\
  \hat e^{(3)}&=&-\frac{2R_gar\sin\theta}{\sqrt{\Sigma A}}dt+\sqrt{\frac{A}{\Sigma}}d\phi
\end{eqnarray}
where $A=(r^2+a^2)^2-a^2\Delta\sin^2\theta$, $\Sigma=r^2+a^2\cos^2\theta$ and $\Delta=r^2-2R_gr+a^2$ \citep{bardeen}.

The Doppler factor of the photon emitted at an angle $\alpha_{em}$ with respect to the direction of the orbital motion in the LNRF is 
\begin{equation}
  \delta_{em}=\frac{\sqrt{1-\left(V^{(\phi)}\right)^2}}{1-V^{(\phi)}\cos(\alpha_{em})}
\end{equation}
Energy of the photons measured by an observer at infinity, $E_{obs}$, is related to the reference line energy $E_{em}$ through the conserved zero component of the photon momentum $p_\mu$. It is expressed in the Boyer-Lindquist coordinates as
\begin{eqnarray}
  \label{eq:Eobs}
  &&E_{obs}=p_t=\hat e^{(a)}_tp_{(a)}\\&& =\delta_{em} E_{em}\left[\sqrt{\frac{\Sigma\Delta}{A}}-\frac{2R_g ar\sin\theta}{\sqrt{\Sigma A}}
  \cos\alpha_{em}\right]\nonumber
\end{eqnarray}
where $p_{(a)}$ is the photon four-momentum in the LNRF. 

In the image of the strongly inclined accretion disk in the left panel of Fig.~\ref{fig:redshift_image} one can clearly see the Doppler effect of the disk rotation, which leads to the red shift of the right part of the disk image, where $\alpha_{em}\approx\pi$, and to the blue shift of the left part of the image, where $\alpha_{em}\approx 0$. From Eq.~(\ref{eq:Eobs}) one can see that the red / blue shift dependence on the angle of emission is through the conventional Doppler factor in the LNRF and additionally through the effect of the LNRF dragging by the black hole rotation. 

The energy profile of the relativistically broadened line can be obtained via a convolution of the maps of $\delta_{obs}=(E_{obs}/E_{em})$ (see Fig.~\ref{fig:redshift_image}) with the radial intensity profile of the line emission $I_{em}(r)$:
\begin{equation}
  F(E_{obs})=\int\int_{\alpha,\beta} \delta_{obs}^3(\alpha,\beta) I_{em}\left(r(\alpha,\beta)\right)d\alpha d\beta
\end{equation}
where $\alpha,\beta$ are the coordinates on the observer's ``screen''. 

\begin{figure}
  \includegraphics[width=\columnwidth]{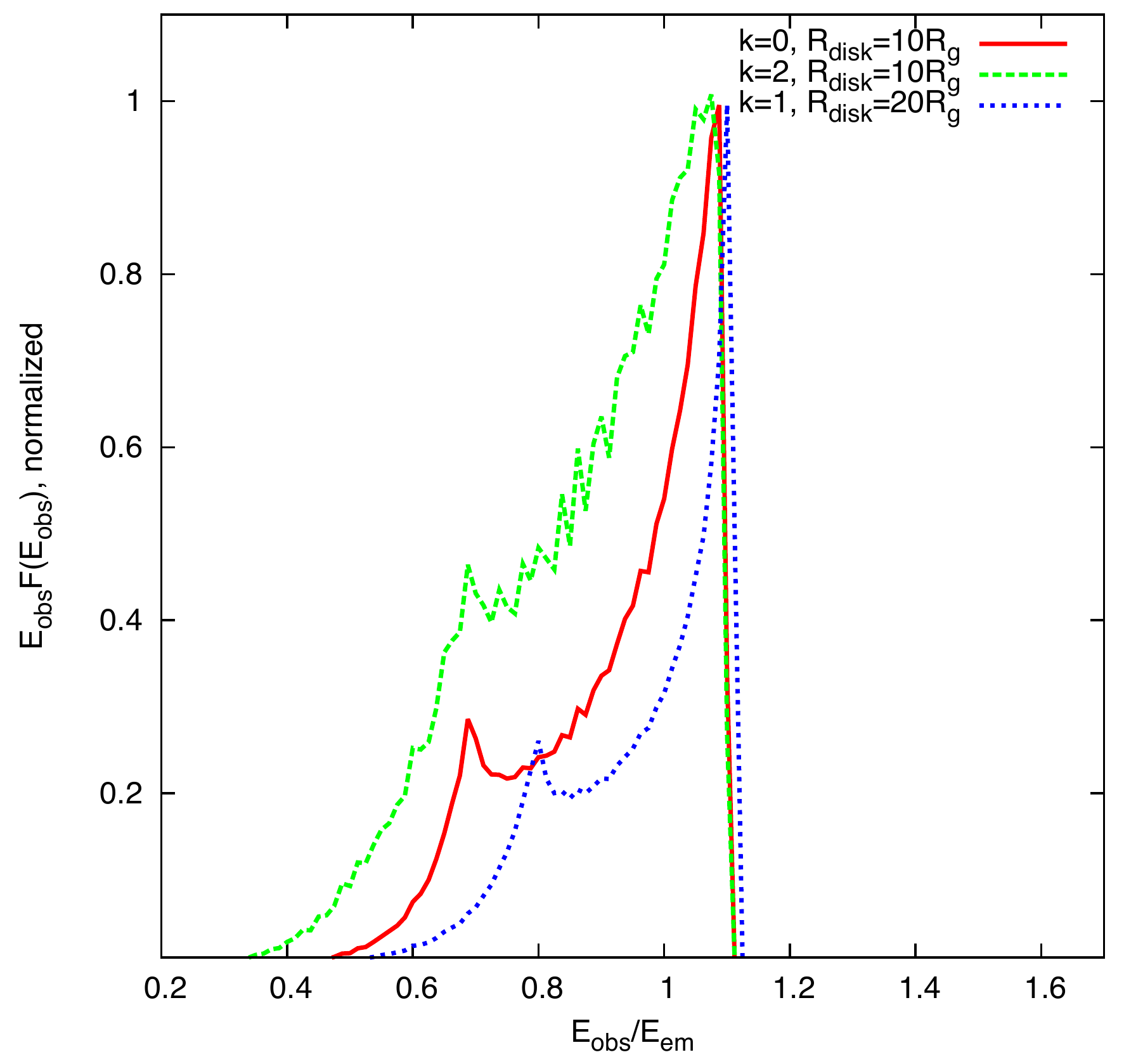}
  \caption{Spectra of the broad emission line from a region of the size $R_{disk}$ with radial brightness profile $r^{-k}$ in the disk around a black hole with specific angular momentum $a=0.9$, inclined at $i=45^\circ$ with respect to the line of sight. }
  \label{fig:line_spectrum}
\end{figure}

Examples of such convolution for different radial profiles $I_{em}\sim r^{-\kappa}$ and different outer radii of the line emission region are shown in Fig.~\ref{fig:line_spectrum}. One can see that the functional shape of the broad line spectrum depends on the parameters like the radial brightness profile and the size of the line emission region. These parameters are not directly measured. Instead, they have to be derived from the observational data. It is clear that model fits to a rather wide range of spectral shapes could be found by the appropriate adjustments of the functional parameter $I_{em}(r)$. 

Thus, even though the spectral shape of the relativistically broadened line is sensitive to the details of the relativistic gravity theory in the strong field regime, measurements of the spectral shape alone can not be used to test the relativistic gravity theory. One of the discussed possibilities for the tests of the strong gravity using the data on the relativistically broadened lines is to complement the line spectrum measurements with the additional timing information, available e.g. in the out-of-steady-state configuration of the accretion disk~\citep{bambi13,johannsen13}. In the following section we propose an alternative type of timing analysis, applicable in the case of the broad line emitting system which is a part of the strong gravitational lens. 

\section{Microlensing of the relativistically broadened lines in strongly lensed quasars }

\subsection{Qualitative picture}

Quasars emitting relativistically broadened emission lines can occasionally be parts of the strong gravitational lens systems~\citep{schneider92,blandford92,claeskens02}, in which case multiple images of the quasar may appear. Typical deflection of light rays by the lensing galaxy of the mass $M_L$ is by an angle
\begin{equation}
  \Theta_E=\sqrt{\frac{4G_NM_L d_{LS}}{d_Ld_S}}\sim 1.5''\left[\frac{M_L}{10^{12}M_\odot}\right]^{1/2}\left[\frac{d}{1\mbox{ Gpc}}\right]^{-1/2}
\end{equation}

In addition to the strong macro-lensing effect, the source images are affected by the effect of microlensing by massive sub-structures in the lensing galaxy~\citep{chang79,kayser86,kochanek04,schmidt10}. The most commonly considered type of microlensing is by stars, i.e. point-like objects with the masses $m\sim M_\odot$. An individual star significantly magnifies the flux from the background sources situated within its Einstein ring
\begin{equation}
  \theta_E=\sqrt{\frac{4G_Nm d_{LS}}{d_Ld_S}}\sim 1.5\left[\frac{m}{M_\odot}\right]^{1/2}\left[\frac{d}{1\mbox{ Gpc}}\right]^{-1/2}\mu\mbox{as}
\end{equation}
Here $d_L,d_S,d_{LS}$ are the angular diameter distances to the lens, to the source and between the lens and the source. $d\sim d_L,d_s,d_{LS}$ is the distance scale. The projected linear size of the Einstein ring on the source plane is
\begin{equation}
  r_E=\sqrt{\frac{4G_Nm d_{LS}d_S}{d_L}}\sim 2\times 10^{16}\left[\frac{m}{M_\odot}\right]^{1/2}\left[\frac{d}{1\mbox{ Gpc}}\right]^{1/2}\mbox{cm}
\end{equation}
The microlensing selectively magnifies the flux from sub-structures with the linear size $R_S$ smaller than $r_E$. In this case the magnification factor can be estimated as
\begin{equation}
  \label{eq:mum}
  \mu_{ml}\sim \left(\frac{r_E}{R_s}\right)^{1/2}
\end{equation}

\begin{figure}
\includegraphics[width=\columnwidth]{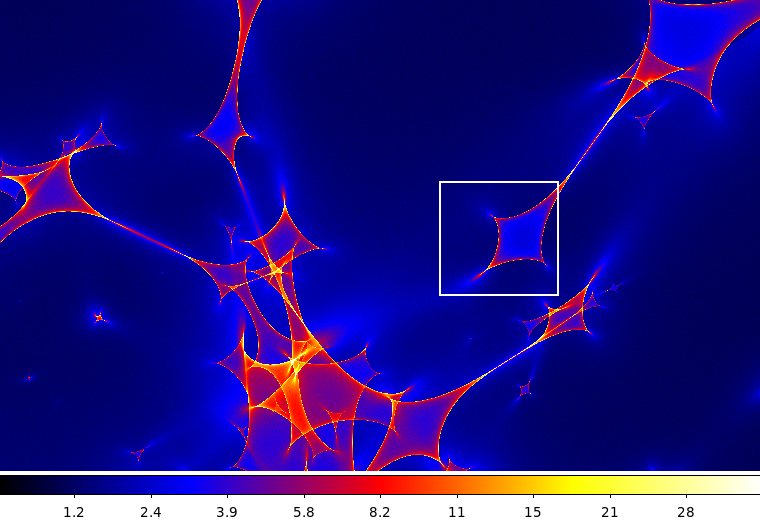}
\caption{Example of the network of the microlensing caustics for the microlensing optical depth $\tau\simeq 0.7$ for the region approximately $2.5 \times 10^{17}$~cm by $1.5 \times 10^{17}$~cm in size. Colour scale represents the magnification. White box marks the region of caustics which we used for the calculations of the effect of caustic crossing on the line emission.}
\label{fig:caustics}
\end{figure}

If the lens galaxy mass in its innermost part within $R_E=d_L\Theta_E\sim 6\left[M_L/10^{12}M_\odot\right]^{1/2}\left[d_L/1\mbox{ Gpc}\right]^{1/2}\mbox{ kpc}$ is dominated by stars, the total number of the micro lenses inside the macro-lensing Einstein ring of the radius $\Theta_E$ is  $N_{ml}\sim M_L/m$. The Einstein circles of individual stars overlap in projection on the sky, because the total area covered by the Einstein rings of the individual stars is comparable to the area of the macro-lensing Einstein ring:
\begin{equation}
  \tau_{ml}=\frac{N_{ml}\pi \theta_E^2}{\pi\Theta_E^2}\simeq 1
\end{equation}
The large, $\tau_{ml} \gtrsim 1$, ``microlensing depth'' leads to a complicated pattern of (de)magnification and to formation of a network of caustics. An example of such a network is shown in Fig. \ref{fig:caustics}, calculated using the ray-tracing method of \cite{wambsganss90,bate10} for $\tau_{ml} = 0.7$ and the initial stellar mass function of~\cite{IMF_Chabrier}. The magnification factor of a point source situated exactly behind the caustic curve is infinite.  The maximal magnification factor for a source of finite size $R_S$ can be estimated from Eq.~\ref{eq:mum}.

The microlensing magnification factor is variable in time, because of the relative motion of the source, of the lens, of stars in the lens galaxy and of the observer. Typical estimates of the relative projected velocity are $v\sim 10^3$~km/s. A source of the size $R_S$ passing behind the caustics stays magnified during a period of time
\begin{equation}
  \label{eq:time}
  t_{ml}\simeq \frac{R_S}{v}\simeq 1\left[\frac{R_S}{3\times 10^{14}\mbox{ cm}}\right]\left[\frac{v}{10^3\mbox{ km/s}}\right]^{-1}\mbox{month}
\end{equation}

The time variability effect of the microlensing is now observed in a range of the strongly lensed quasars, via the measurement of the difference in the lightcurves of the multiple images of the lensed sources~\citep{chang79,schmidt10,dai10,chen12,chartas12,macleod15}. Measurements of the strength of the microlensing effect leads to the estimates of the sizes of the regions of visible~\citep{kochanek04,eigenbrod08}, and X-ray~\citep{dai10,chen12,chartas12,macleod15}  emission in the lensed quasars. 

\begin{figure*}
  \includegraphics[width=\linewidth]{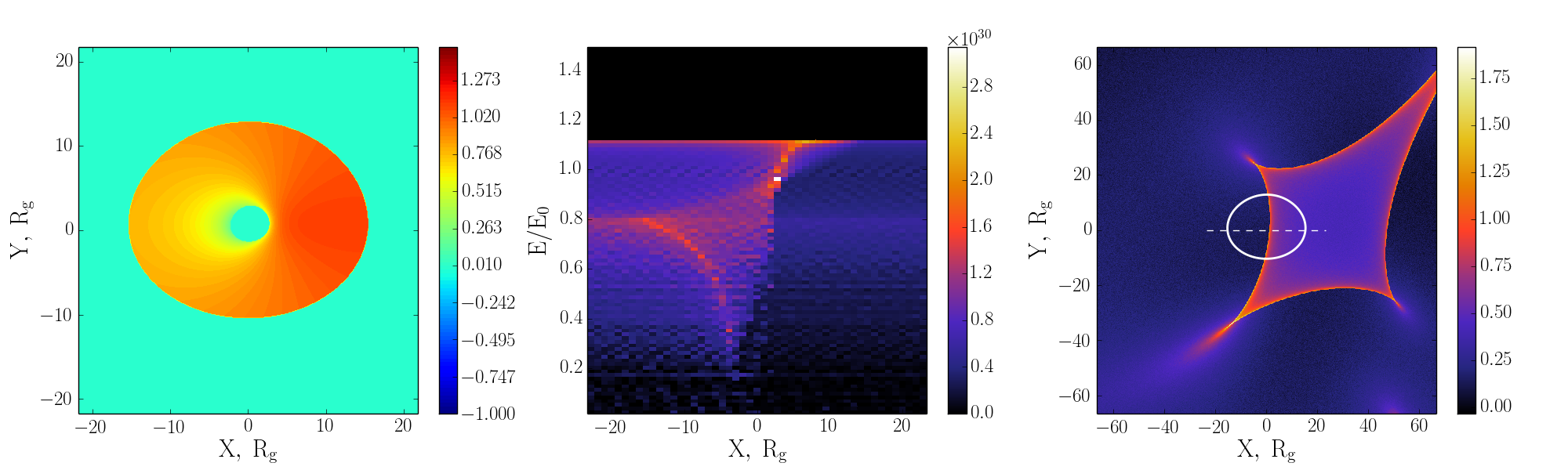}
  \includegraphics[width=\linewidth]{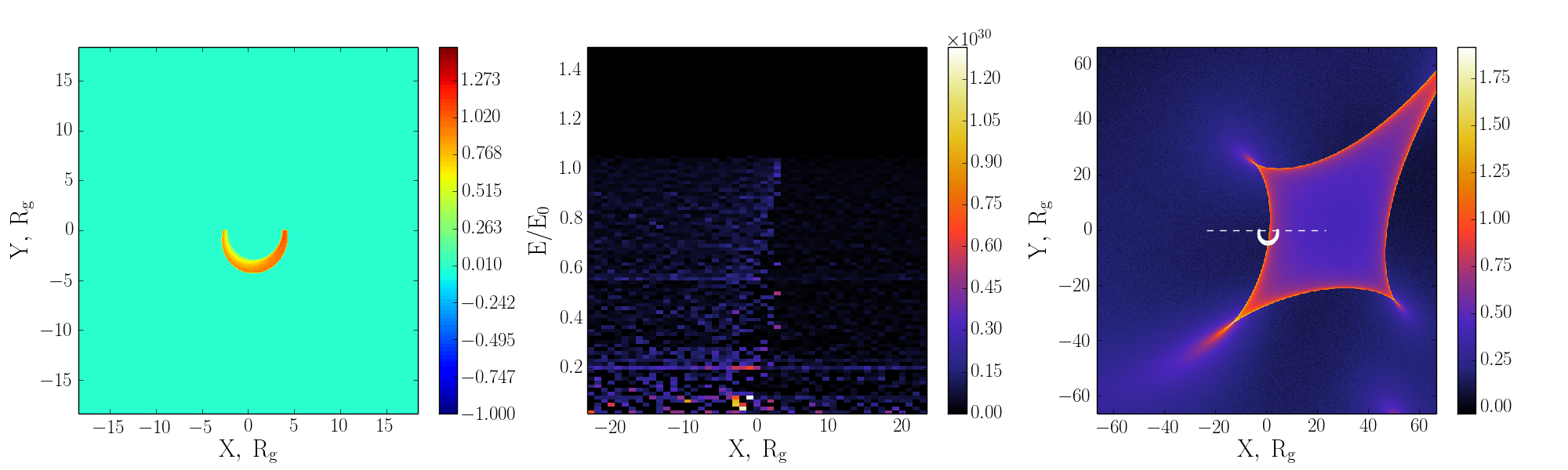}
  \caption{
  \textit{Top left:} image of the red / blue shifted emission from an accretion disk with external radius $20R_g$ around a black hole with specific rotation moment $a=0.9R_g$ inclined at $i=45^\circ$ with respect to the line of sight. Color scale gives the red / blue shift $E_{obs}/E_{em}$. 
  \textit{Top right:} a microlensing caustic and its trajectory (dashed line) in front of the accretion disk image (shown by a solid contour). Colour scale represents the $\log_{10}$ of the microlensing magnification.
  \textit{Top middle:} evolution of the spectrum of relativistically broadened line during the period of caustic crossing in front of the disk image. $x$ axis is the distance of the point of intersection of caustic with the line of nodes of the disk, measured in the units of $R_g$. Only first order image is taken into account. Color scale gives the relative intensity in arbitrary units.
  \textit{Bottom:} the same as the top images, but for the second order image of the disk.
  }
  \label{fig:profile_a01_view80}
\end{figure*}

\subsection{``Microlensing portrait'' of the relativistically broadened line}

If accretion disk is a source of the relativistically broadened line emission,  the line flux is also subject to the microlensing effect. Strong magnification of the line flux is expected during the periods of caustics crossings of the disk images~\citep{popovic06,chartas12}. 

In general, the caustics crossing does not affect equally the flux of the ``red'' and ``blue'' wing of the line.  From Fig.~\ref{fig:redshift_image} one could guess that the caustic, passing in front of the disk image, say from right to left would first magnify the flux of the red wing of the line and then the flux of the blue wing of the line (the only exception is when the caustic is parallel to the line of nodes of the disk). This expectation is confirmed with the numerical simulation shown in Fig.~\ref{fig:profile_a01_view80}. The left panel of the figure shows an accretion disk inclined at $i=45^\circ$ and truncated at the radius $r=20G_NM$. The disk surface brightness profile is assumed to be $I(r)\propto r^{-2}$. The right panel of the figure shows the trajectory of the disk passing behind a caustic taken from the map of Fig.~\ref{fig:caustics}. The motion of the disk is along the dashed line, so that the caustic passes perpendicularly to the line of nodes of the disk.

The middle panel of Fig.~\ref{fig:profile_a01_view80} shows the evolution of the spectrum of the relativistically broadened line during the period of caustic crossing. The $y$ axis shows the energy in the units of the reference line energy. The $x$ axis shows the position of the caustic centre with respect to the black hole and effectively represents the time. The colour scale shows the intensity of emission at different energies and different moments of time. 
One can see that the amplification of the flux from the part of the disk which is exactly behind the caustic leads to the appearance of an increased flux between some minimal and maximal energy. The increased flux component has pronounced red and blue ``edges''. The energy of the ``edge'' is determined by the energy of photons $E_{obs}$~(Eq.~\ref{eq:Eobs}), calculated for the photons emitted from the
\begin{itemize}
  \item point of intersection of the caustic with the line of nodes of the disk, if the caustic does not pass through the ISCO or
  \item points of intersection of the caustic with the ISCO, if the caustic is in front of the ISCO.
\end{itemize}
The energy of the edge evolves with time because the points of intersection of the caustic with the line of nodes of the disk and with the ISCO move as the caustic advances across the disk image. 

\begin{figure}
\includegraphics[width=\linewidth]{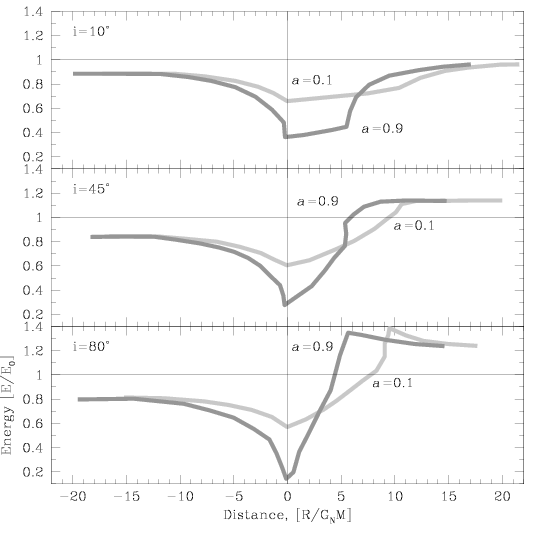}
\caption{Evolution of the energy of the red / blue edges of the line throughout the caustic crossing period for different rotation momenta of the black hole and different viewing angles. }
\label{fig:portrait_perpendicular}
\end{figure}

In principle all the ${\cal N}$th order lensed images of the disk contribute to the line flux. However, the flux from higher order images is typically weaker than that of the first order image. The bottom panel of Fig. \ref{fig:profile_a01_view80} shows the evolution of the line spectrum produced by the microlensing of the second order image. One could see that the evolution of the second order spectrum is significantly different from that of the first order image. If the sensitivity of telescope is sufficient, the spectral evolution should reveal multiple edges corresponding to the images of different order.

The shape of the curve which the red and blue edge draw in the middle panel of Fig.~\ref{fig:profile_a01_view80} depends on two parameters: the rotation moment of the black hole $a$ and the inclination angle of the disk $i$.  The shape does not depend on numerous additional parameters of the physical system harbouring the black hole, the accretion disk and the X-ray source illuminating the accretion disk. It also does not depend on the parameters of the microlensing caustics, such as the projected velocity of the caustic, the angle of inclination of the caustics with respect to the line of nodes of the disk, the typical mass of the stars in the lensing galaxy and their distribution in masses etc. It is rather the time dependence of the full spectrum of the line which is sensitive to the numerous astrophysical parameters. A range of the edge time evolution curves for different values of $a$ and $i$ parameters is shown in Fig.~\ref{fig:portrait_perpendicular}. In a sense, these red / blue edge time evolution curves could be called the ``microlensing portraits'' of the broad emission lines.

The independence of the shape of the microlensing portrait of the line on the system parameters could be understood in simple terms if one uses a toy model of the accretion disk based on the Newtonian gravity instead of the GR (or an alternative theory of gravity). This toy model neglects the gravitational redshift of photons, taking into account only the Doppler effect. 
The circular orbits in Newtonian gravity have velocities 
\begin{equation}
  v_K=\sqrt{\frac{G_NM}{r}}
\end{equation}
The maximal Doppler red / blue shift of line emission produced at the distance $r$ occurs for the emission produced at the line of nodes of the disk. It is determined by the inclination angle $i$ of the disk
\begin{equation}
  g(r) = v_K\cos(i) = \sqrt{\frac{G_NM\cos^2i}{r}}
\end{equation}

The microlensing caustic is a line inclined at an angle $\alpha_c$ with respect to the line of nodes of the disk. If the caustic moves with the projected velocity $v_c$, the motion of the intersection of the caustic with the line of nodes is described by the relation 
\begin{equation}
  \label{eq:rt}
  r(t)=v_ct\cos\alpha_c
\end{equation}
so that the time evolution of the maximal / minimal blue / red shift is 
\begin{equation}
g(t)=\pm \sqrt{\frac{G_NM\cos^2i}{v_c\cos\alpha}}\sqrt{\frac{1}{t}}
\end{equation}
The functional shape of $g(t)$ is $1/\sqrt{t}$ and it does not depend on the parameters of the system. In fact, in non-relativistic theory the functional shape of the curve is determined by the gravity theory itself, rather than by the parameters of the system. 
If the gravity force law would deviate from the $1/r^2$, the functional shape of the $g(t)$ curve would deviate from $1/\sqrt{t}$. 

Variations of parameters of the system do not change the shape of the curve, but instead they influence the scaling of the $x$ and $y$ axis in the ``microlensing portrait'' of the curve. For example, changing the velocity of motion of the caustic would lead to stretching or squeezing of the time scale of the caustic crossing, i.e. the scale of the $x$ axis. A similar effect occurs if the angle $\alpha_c$ changes. Only the projected velocity of the caustic, $v_c\cos\alpha$, matters. Increase or decrease of the mass of the black hole also changes the scaling of the $x$ axis, because the natural scale of variations of the red / blue shift is given by the gravitational radius $R_g=G_NM$. 

Description of the ``microlensing portrait'' of the line within GR introduces only technical complications, but does not change this qualitative picture. The red/blue shift at the point of intersection of caustic with the line of nodes or with the ISCO can be calculated from Eq.~(\ref{eq:Eobs}), $g=E_{obs}/E_{em}-1$.

The parameters $r, \alpha_{em}$ in (\ref{eq:Eobs}), which are functions of time,  have to be calculated numerically via back-tracing of photons from infinity to their emission points in the disk. The asymptotic form of these functions for the times when the intersection of the caustic with the line of nodes of the disk is far away from the black hole is given by $\alpha_{em}\simeq \pi/2-i$ and by Eq. (\ref{eq:rt}).  

The redshift $g$ has two contributions. In addition to the Doppler effect given by the Doppler factor $\delta_{em}$, the line emission is gravitationally redshifted. The functional shapes of both the gravitational redshift and of the Keplerian velocity depend on the black hole rotation momentum $a$. 

One more important difference between non-relativistic and relativistic gravity model is that in non-relativistic gravity the function $g(t)$ is not bounded from above and from below. This is because the circular orbits can be arbitrarily close to the centre of gravity. To the contrary, in GR the stable circular orbits exist only down to the distance of the ISCO. Finite gravitational and Doppler red/blue shifts of photons emitted from the ISCO explain the ``plateau'' parts of the ``microlensing portraits'' of the line shown in Fig.~\ref{fig:portrait_perpendicular}.

\subsection{Celestial mechanics of circular orbits around black hole from the microlensing portraits of emission lines}

Fig.~\ref{fig:portrait_perpendicular} shows that the edge evolution curves (the microlensing portraits) form a family of curves parametrised with the black hole rotation moment $a$ and the disk inclination angle $i$. The scaling of the $x$ and $y$ axes depend on two additional parameters: the angle between the caustic and the line of nodes of the disk $\alpha$ and the velocity $v$, which provides the relation $x=vt$ between the time $t$ and the $x$ coordinate in Fig.~\ref{fig:portrait_perpendicular}. Overall, the full curves depend on four parameters. These four-parameter curves could be fitted to the observational data, presented in the same form as  the simulated data shown in the middle panel of the Fig. \ref{fig:profile_a01_view80}. Note that in spite of the fact that the overall evolution of the form of the spectrum of the broad line throughout the caustic crossing period is quite complicated, the evolution of the edge could be determined rather reliably and independently from the details of the complex evolution of the spectrum. Taking into account sharpness of the edge, one could use the dedicated imaging analysis methods (e.g. wavelet analysis) to single out the edge evolution on top of the overall spectral evolution.

If the quality of the data is high enough, fitting of the model curves to the data provides a test of the relativistic gravity theory (if the number of data points in the edge evolution curve is larger than four). If the sensitivity of the telescope is high enough, even more constraints on the gravity theory could be derived from the observation of the edge evolution curves of the higher order images (see the bottom panel of Fig. \ref{fig:profile_a01_view80}), because the shape of these curves does not depend on any additional parameters.

Depending on the orientation of the accretion disk with respect to the line of sight, the edge evolution curves provide different types of information on the details of the structure of gravitational field and of the motion of massive test bodies and photons in this gravitational field. 

If the disk is observed almost face-on (the upper panel of Fig. \ref{fig:portrait_perpendicular}), the edge evolution curve has two red branches ($E_{obs}/E_{em}<1$)  and no blue branch. This is obviously due to the effect of the gravitational redshift. The asymptotic behaviour of the two red branches at large $x=vt$  is almost identical. Taking the moment of time with respect to which the two edges are symmetric, one could define the moment when the caustics crosses the centre of the disk image. If the time is counted with this moment taken as a zero reference time, the shape of the $g(t)$ curve provides a measurement of the gravitational redshift as a function of the distance from the black hole.

The ``plateau'' of the edge evolution curve, visible in the upper panel of Fig.~\ref{fig:portrait_perpendicular}, corresponds to the time interval when the caustic passes the ISCO. The depth of the plateau is determined by the gravitational redshift of photons emitted almost perpendicularly to the disk from the ISCO. This redshift depends on the black hole rotation moment $a$. The width of the plateau is determined by the size of the ISCO which is also a function of $a$.  

If the disk is significantly inclined, the Doppler effect ``tilts'' the microlensing portrait, lifting one side of the edge curve up and pushing the other side down energy-wise. Obviously, the measurement of the distance (time) dependent tilt of the curve provides a measurement of velocities of the circular orbits as a function of the distance.

If the black hole rotation moment $a$ is not zero, the inclination of the disk not only ``tilts'' the plateau corresponding to the period of passage of the ISCO, it also deforms its shape. This is because  of the fact that even in the absence of the Doppler effect due to the disk rotation, there is an angular dependence of the gravitational redshift in Eq.~(\ref{eq:Eobs}) caused by the rotational drag by the black hole. Photons reaching the observer from different points of the ISCO are emitted in different directions spanning a cone with the axis given by the black hole rotation axis and with an opening angle comparable to the inclination angle of the disk. Evolution of the energy of the red / blue edge of the line during the period of caustic crossing of the ISCO provides a ``scan'' of the red / blue shift of the photon energy as a function of direction along the cone. 
 
If the quality of the spectral measurements is high enough, the angle between the caustics and the line of nodes of the disk could be measured from an effect of ``splitting'' of the edge during the period of caustic passage of the ISCO. The origin of such splitting is clear from  Fig.~\ref{fig:redshift_image}. The red / blue shifts  of photons emitted from the points of intersection of the caustic with the ISCO are different. If the caustic is perpendicular to the line of nodes, this difference is small and the effect of the splitting of the edge is difficult to observe. However, if the caustics is strongly inclined, the effect of the splitting by the caustic is more pronounced and could be used for the measurement of the angle $\alpha$.

\section{Discussion}

The results of the previous section show that measurements of the evolution of the red / blue edge of line emission from the accretion disk during the periods of crossing of the disk image by the microlensing caustic provide a unique possibility to make a range of measurements, which could be used to characterise the structure of the strong gravitational field in the direct vicinity of the black hole horizon.

The functional shape of the microlensing portraits of the lines, shown in Fig.~\ref{fig:portrait_perpendicular}, is largely independent from the parameters of the accretion disk and of the caustic passing in front of the disk image. This near-parameter-independence indicates that the measurements of the ``microlensing portraits'' of the lines could potentially be used to test the GR in the strong field regime near the black hole horizon. Measurements of the ``microlensing portraits'' would, in a sense, be equivalent to the measurements leading to the Kepler laws, on the basis of which the celestial mechanics of the Solar system was established and the non-relativistic gravity law was derived in the past.

The most straightforward observational example of the relativistically broadened line is the Fe K$\alpha$ line at the reference energy 6.4~keV~\citep{fabian00}. Broad Fe~K$\alpha$ line is observed in a large number of Seyfert galaxies and quasars, starting from the initial ASCA discovery of the line in MCG~15-6-30 galaxy spectrum~\citep{tanaka95}. 
 
A number of strongly lensed quasars, such as e.g. \rxj, possess the broad Fe~K$\alpha$ line in their spectra~\citep{chartas12,reis14}. Moreover,~\cite{chartas12} find that the broad Fe~K$\alpha$ line emission of \rxj\ is influenced by the microlensing. They interpret the shifts in the energy of the peak of the line as the magnification due to the microlensing effect. Unfortunately, the time coverage of the observational campaign reported in~\cite{chartas12} was not sufficient for the measurement of the evolution of the line (or edge) energy with time. An estimate of the black hole mass in \rxj\  is $M\sim 10^8M_\odot$, so that the caustic crossing takes approximately one month (see Eq.~\ref{eq:time}). This time was also the typical time interval between the subsequent observations in the observational campaign of the source reported by~\cite{chartas12}. Only two observations are available for the entire caustic crossing period.

In spite of the fact that the observations of \rxj\ do not allow the measurement of the ``microlensing portrait'' of the Fe~K$\alpha$ line, they are useful for the demonstration of the feasibility of such measurement. The shift of the line energy due to the microlensing was observed in the weakest component ``D'' of the source image. Brighter images are also known to be affected by the microlensing variability~\citep{dai10}. Monitoring of the source in X-rays should sooner or later catch the caustics crossing periods in the brighter components of the source image. 

The caustics crossings of the image components of the \rxj\ and of other strongly lensed quasars should be frequent enough (an estimate of ``several'' per decade could be derived for \rxj\ from~\cite{dai10}). An efficient observational approach for the measurement of the ``microlensing portrait'' of the Fe~K$\alpha$ line should include also an efficient ``forecasting'' of the caustics crossing periods, followed by the intensive observational campaign with the dense time coverage in X-rays, upon the positive outcome of the forecast of the caustic crossing period. 

The forecasting of caustics crossings could be done using the source monitoring campaigns in X-rays or in the visible band. The overall size of the X-ray emitting part of the disk is just about 10 gravitational radii. It is typically much smaller than the Einstein radius of the microlensing centres. This results in strong magnification of the X-ray flux of individual images of the source, by up to one-two orders of magnitude during the periods of the caustics crossings.

Such strong increase of the flux of one of the source images could be detected via a monitoring campaign of a strongly lensed source with an X-ray telescope with even a moderate angular resolution, which is not able to resolve the multiple images of the source (like e.g. the Swift/XRT instrument). From the pattern of the magnification maps in Fig.~\ref{fig:caustics} one can see that the caustic crossing moment should be preceded by a period of the steep rise of the magnification factor in about 50\% of caustics crossing episodes due to the presence of the ``plateau'' regions in the caustic shapes. Identification of such steep flux rises in caustic crossing precursors with a regular monitoring campaign of a lensed source (year-to-multi-year long) should pose no problem.

The size of the accretion disk in the visible band is significantly larger than the gravitational radius~\cite[by one-two orders of magnitude,][]{dai10,eigenbrod08,macleod15}. The overall changes of magnitudes of the source images due to the microlensing are moderate, because of the large source size. Still, the caustic crossing periods should be identifiable as the periods of the rapidly growing amplitude of the difference of the flux of one of the source images. The best possibility for the caustics crossing detection is, perhaps, a combination of the data of the visible and X-ray band monitoring. Optical monitoring of the source could serve for the confirmation of the approach of the caustic crossing moment, identified based on the detection of the ``X-ray precursor''.

As soon as the caustic crossing precursor is detected, an intense observational campaign with more powerful X-ray telescopes which are able to resolve the individual source image components could be initiated, with the main goal to obtain a high signal-to-noise ratio ``microlensing portrait'' of the relativistically broadened line, like the one shown in the middle panel of Fig.~\ref{fig:portrait_perpendicular}.

The quality of the test of GR, which could be obtained via the measurement of the microlensing portrait of the Fe~K$\alpha$, line depends on the precision of the measurement of the edge evolution curve. This precision depends on the number of the observations, carried out throughout the caustic crossing period, and on the precision of the measurement of the edge position in each observation. Minimum five observations per caustic crossing are required to fully determine the parameters of the curve. Each additional data point is potentially useful for the test of the theory. 

First constraints on the relativistic gravity theory could be obtained using the proposed method already with the current generation X-ray telescopes -- Swift/XRT, XMM-Newton, Chandra and Suzaku. Higher precision of the test can be obtained with the high spectral resolution measurement, e.g. the X-ray spectrometers on Astro-H~\citep{Astro-H} and/or Athena~\citep{Athena}. 


\end{document}